\documentclass[useAMS,usenatbib]{mn2e}

\usepackage{graphicx}
\def \aap{A\&A}
\def \aapr{A\&AR}
\def \aj{AJ}
\def \apj{ApJ}
\def \apjl{ApJL}
\def \mnras{MNRAS}
\def \nat{Nature}
\def \pasp{PASP}
\def \procspie{Proceedings of the SPIE}

\title[Spotting the misaligned outflows in NGC~1068 using X-ray polarimetry]{Spotting the misaligned outflows in NGC~1068 using X-ray polarimetry}
\author[R.~W.~Goosmann and G.~Matt]{R.~W.~Goosmann$^{1}$\thanks{E-mail: rene.goosmann@astro.unistra.fr (RWG)} 
                                    and G. Matt$^{2}$\thanks{E-mail: matt@fis.uniroma3.it (GM)}\\
$^{1}$Observatoire astronomique de Strasbourg, Equipe Hautes Energies, 11~rue de l'Universit\'e, F-67000 Strasbourg, France\\
$^{2}$Dipartimento di Fisica, Universit\`a degli Studi Roma Tre, Via della Vasca Navale 84, I-00146 Roma, Italy}

\begin{document}

\date{Accepted 2011 April 15. Received 2011 April 11; in original form 2010 December 21}

\pagerange{\pageref{firstpage}--\pageref{lastpage}} \pubyear{2010}

\maketitle

\label{firstpage}

\begin{abstract}
We model the expected X-ray polarisation induced by complex reprocessing in the active nucleus of the Seyfert-2 galaxy NGC~1068. Recent analysis of infrared interferometry observations suggests that the ionised outflows ejected by the central engine are not aligned with the symmetry axis of the obscuring torus. This conclusion was obtained by extrapolating the apparent orientation of the narrow line region to the inner parts of the ionisation cones. We show that future measurements of the soft X-ray polarisation vector unambiguously determine the orientation of the ionisation cones. Furthermore, X-ray polarimetry across a broad photon energy range may independently verify the misalignment between the ionisation cones and the axis of the torus. To model the expected polarisation percentage and position angle, we apply the radiative transfer code {\sc stokes}. Reprocessing of the primary X-ray radiation takes place in the accretion disc, the surrounding equatorial torus and the inclined, ionised outflows. We also examine additional equatorial scattering occurring in between the accretion disc and the inner surfaces of the torus. Radiative coupling between the different reprocessing components is computed coherently. The resulting polarisation properties depend on the optical depth of the reprocessing regions and on the viewing angle of the observer. We show that even under unfavourable conditions the misalignment of the outflows with respect to the torus axis can be determined from a rotation of the polarisation position angle between softer and harder X-rays. We argue that the misalignment of the outflows with respect to the torus axis in NGC~1068 may be constrained by a future X-ray mission if equipped with a broad band polarimeter.
\end{abstract}


\section{Introduction}
\label{sec:intro}

The standard unified scheme for active galactic nuclei (AGN) attempts to explain the observational appearance of different AGN types mainly as an inclination effect \citep{antonucci1993,urry1995}. The continuum emission being produced by the accretion disc of the central supermassive black hole (SMBH) irradiates the surrounding broad line region (BLR) and gives rise to broad optical/UV line emission. Obscuring equatorial dust at larger distances plays a crucial role in the unified model as at higher inclinations it covers the BLR and thereby disentangles type-1 AGN, which spectroscopically show broad line emission, from type-2 objects, which do not. The accretion process onto the SMBH partly leads to re-ejection of matter and supposedly causes ionised winds in the polar direction. The presence of these winds has brought strong support to the unified model as it allowed to indirectly detect hidden BLRs in obscured type-2 objects by scattering of BLR light around the torus \citep{antonucci1985}. Systematic studies in optical spectropolarimetry lead to a classification of Seyfert galaxies according to the direction of the optical polarisation angle \citep[see e.g.][and references therein]{antonucci1984,smith2002,smith2004} that can be explained in the framework of the unified scheme when including equatorial scattering inside the torus funnel.

The simplest approach of the unified model assumes that the rotation axis of the accretion disc is aligned to the axis of the torus and of the outflows. The alignment of these components is justified by assuming a symmetric mass transfer from the inner torus towards the outer accretion disc as well as by the symmetric collimation effect that the torus should have on the polar outflows. But recent work by \citet{raban2009} surprisingly suggests that the outflows of the well-studied Seyfert-2 galaxy NGC~1068 are inclined by $18\degr$ with respect to the apparent axis of the torus \citep[see Fig.~9 in][]{raban2009}. If this result is confirmed it has important consequences for our understanding of the accretion and ejection mechanisms close to active SMBHs. 

In this work, we explore to which extend the misalignment of the outflows with respect to the torus in NGC~1068 can be probed by X-ray polarimetry observations. With the advent of a new generation of X-ray polarimeters based on photo-electric effects, X-ray imaging polarimetry becomes feasible across the 2--35~keV energy range \citep{soffitta2010}. The X-ray emission from AGN is much less sensitive to stellar contributions from the host galaxy than optical/UV radiation is, and therefore it offers a better probe of the reprocessing geometry in the innermost regions of NGC~1068. One could expect that the radiative coupling of centrally emitted X-rays to the torus and the outflows is somewhat similar to the observed behaviour in the optical/UV as in both wavebands electron scattering plays an important role. But in the hard X-ray band electron scattering becomes wavelength-dependent, which should have an impact on the polarisation.

In the following, we build up a sophisticated X-ray reprocessing model of NGC~1068. We start out with a simple irradiated slab representing the accretion disc and an elevated primary source. Then, we add dust obscuration in a torus geometry testing for different optical depths. Next, we include inclined polar outflows and conduct calculations for different optical depths of these ionisation cones. In the soft X-ray band with photon energies $E < 10$~keV absorption processes in the torus are very important and at high inclination angles the primary source and reflection from the accretion disc should only be seen indirectly by scattering in the outflows. The scattering in the ionisation cones should then imprint a polarisation position angle that is perpendicular to the axis of the outflow. At higher energies, when scattering off the torus becomes more prominent, the position angle should change and be determined by the torus geometry. This switch of the polarisation angle depends on the relative Stokes fluxes coming from the ionisation cones and from the torus. Therefore, it must vary with the optical depths in both components as well as with the inclination angle. Finally, we also examine the impact of additional equatorial scattering in the farther away components of the accretion flow that are located between the inner surfaces of the torus and the outer accretion disk. Such an additional scattering component produces polarised reprocessing by itself but also changes the irradiation geometry of the inner surfaces of the torus. 

We test the resulting spectra and polarisation properties for different optical depths of the obscuring dust and of the scattering electrons in the ionisation cones. Aside from the optical depths, our model comprises other interesting parameters such as the half-opening angle of the torus or of the outflows. Their impact is going to be examined elsewhere. This study is particularly designed to predict the expected X-ray polarisation signal if the geometry inferred by \citet{raban2009} is correct. We therefore fix all dimensions and angles according to their findings.

The radiative transfer code applied here is presented in Sect.~\ref{sec:tools} and the details of the model geometry are described in Sect.~\ref{sec:geom}. The results for different combinations of the individual reprocessing regions and then for the case of NGC~1068 follow in Sect.~\ref{sec:results}. We close with a discussion in Sect.~\ref{sec:discuss} where we also shed light on the prospects for future X-ray polarimetry observations of NGC~1068.

\section{X-ray modelling with STOKES}
\label{sec:tools}

We model the radiative transfer and the polarisation signature of the complex X-ray reprocessing environment in AGN by applying the code {\sc stokes} \citep{goosmann2007}. It allows to define rather general emission and reprocessing regions and coherently computes the 3D radiative coupling between them. The model space can then be evaluated from any viewing direction as a function of the photon energy, $E$, in terms of the X-ray spectral flux, $F/F_{*}$, the polarisation degree, $P$, and the polarisation position angle, $\psi$. Note that the flux is normalised to the total flux of the primary source, $F_{*}$, emitted into the same viewing direction. The {\sc stokes} code was originally developed for the optical/UV spectral range and has recently been extended into the X-ray band \citep{goosmann2009}. We generalised the algorithm for electron scattering to include the inelastic Compton effect and we included routines to treat photo-ionisation and recombination effects. Photo-absorption above the iron K-shell and the subsequent emission of iron K$\alpha$ or K$\beta$ line photons is included and weighted against the probability of Auger effects. Throughout this paper, we consider X-ray reprocessing by neutral matter consisting of hydrogen, helium, carbon, nitrogen, oxygen, neon, magnesium, silicate, sulfur and iron with cosmic abundances. 

The X-ray part of the code was tested against previous results of spectral modelling, in particular the recent X-ray reprocessing model for AGN tori presented by \citet{murphy2009} and disc reflection models by \citet{george1991}. Polarisation results were compared against the results given by Ghisellini, Haardt \& Matt (1994).

\begin{table*}
 \centering
 \begin{minipage}{15cm}
  \caption{Characteristics and dimensions of the reprocessing elements used in our modelling. \label{tab:models}}
  \begin{tabular}{@{}ll@{}}
  \hline
  reprocessing region & description\\
  \hline

  central, irradiated slab     &  cylindrical disc with radius of $0.004$~pc\\
                               &  and half-height of $3.25 \times 10^{-7}$~pc,\\
                               &  irradiated by two isotropic sources at\\
                               &  $\pm 0.001$~pc from the disc mid-plane,\\
                               &  filled with optically-thick, neutral matter\\
                               &  \\

  equatorial, obscuring torus  &  torus with elliptical cross-section and half-opening angle $\theta_{\rm tor} = 60\degr$,\\
                               &  inner-radius of $0.1$~pc and outer radius of $0.5$~pc,\\
                               &  filled with neutral matter of an equatorial, hydrogen-equivalent density $N_{\rm H,tor}$\\
                               &  \\

  tilted, conical outflows     &  double-cone with the half-opening angle $\theta_{\rm cone} = 40\degr$,\\
                               &  titled by the angle $\theta_{\rm tilt} = 18\degr$ from the symmetry axis,\\
                               &  inner-radius of $0.3$~pc and outer radius of $1.8$~pc,\\
                               &  filled with electrons of a radial Thomson optical depth $\tau_{\rm cone}$\\
                               &  \\

  equatorial, scattering disc  &  horizontal, flared disc with half-opening angle $\theta_{\rm wedge} = 20\degr$,\\
                               &  with $\theta_{\rm wedge}$ being measured from the equatorial plane,\\
                               &  inner-radius of $0.03$~pc and outer radius of $0.05$~pc,\\
                               &  filled with electrons of a radial Thomson optical depth $\tau_{\rm wedge} = 1$\\
  \hline
  \end{tabular}
 \end{minipage}
\end{table*}

\section{X-ray reprocessing in NGC~1068}
\label{sec:geom}

The modelling geometry we consider is based on several elements. We approximate the reprocessing of an AGN accretion disc \citep{george1991,matt1991,matt1993} by defining a flat, cylindrical, optically thick and neutral disc that is being irradiated from both sides by point-like X-ray sources isotropically emitting a power-law spectrum with a spectral flux $F_{*}(E) \propto E^{-\alpha}$ and $\alpha = 1$. Note that this spectral index $\alpha$ corresponds to a photon index $\Gamma = 2$. This choice of the spectral index is motivated by the analysis of an {\it XMM-Newton} spectrum of NGC~1068 presented in \citet{matt2004}. The two primary sources in our model are located on opposite sides of the disc symmetry axis such that the disc subtends a half-opening angle of $\sim 75\degr$ with the sources. The reprocessing medium thus catches more than 95 per cent of the primary radiation emitted towards the disc. 

\begin{figure}
  \centering
  \includegraphics[width=8.2cm]{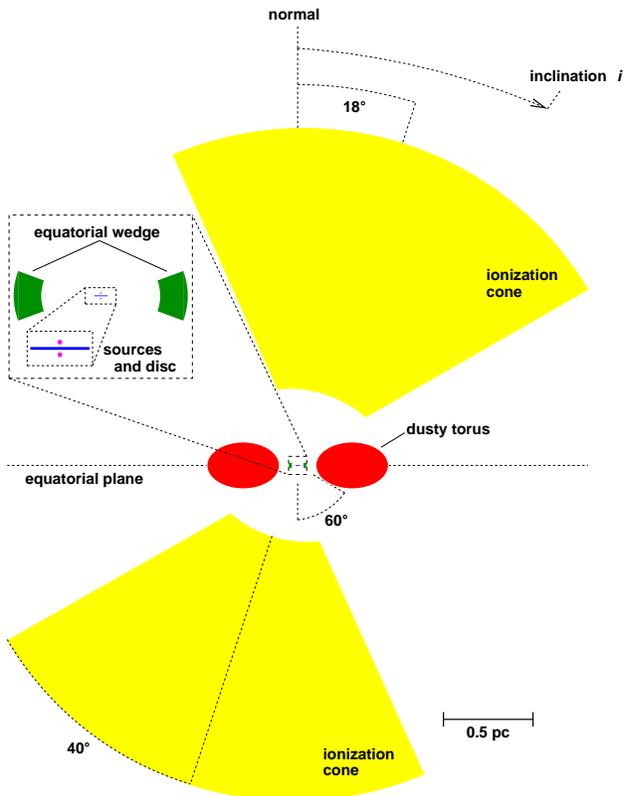}
  \caption{Sketch of the emission and scattering geometry in this modelling. The insets show a consecutive zoom on the central elements representing a neutral slab that is surrounded by an equatorial scattering wedge and irradiated on both sides by point-like, primary X-ray sources.}
  \label{fig:sketch}
\end{figure}

To this base element of our modelling scenario we then add up to three more reprocessing regions: an equatorial, obscuring torus with an elliptically-shaped cross-section (1), tilted, conical outflows in the polar directions (2), and a ring-like scattering region in the equatorial plane, located between the accretion disk and the inner surfaces of the torus (3). The equatorial torus is filled with neutral matter and centred on the much smaller reprocessing disc. The dimensions of all emission and scattering geometries are summarised in Table~\ref{tab:models} and a sketch is provided in Fig.~\ref{fig:sketch}. The symmetry axes of the torus, the equatorial wedge and the disc are identical and the torus funnel has a half-opening angle $\theta_{\rm tor} = 60\degr$, as taken from \citet{raban2009}. Thus, the direct view of the central components is obscured at viewing directions, $i$, with $i > \theta_{\rm tor}$. Both angles are measured with respect to the symmetry axis.

We first only consider reprocessing off the irradiated accretion disc alone and then add the torus. The following scenarios include polar outflows that we realise by a tilted double-cone. The centre of the cone is identical with the ones of the reprocessing disc and the torus. The double-cone half-opening angle is measured by $\theta_{\rm cone} = 40\degr$ and its tilting angle with respect to the torus symmetry axis by $\theta_{\rm tilt} = 18\degr$ \citep[see][]{raban2009}. We assume that the matter inside the double-cone is ionised and dominated by electron scattering. The Thomson optical depth, $\tau_{\rm cone}$, is measured between the inner and outer radii of the cone. The misalignment between the torus and the ionisation cones is inferred from the mid-IR interferometry measurements of \citet{raban2009} in combination with the maser observations of \citet{gallimore1996} as well as with the {\it HST}-spectroscopy modelling conducted by \citet{das2006}. For further details on the geometrical considerations we refer to the very instructive discussion given in \citet{raban2009}. 

There is another, potentially important scattering region that is required to explain the dichotomy of the optical/UV polarisation angle. The equatorial scattering ring as described by \citet{antonucci1984} and \citet{smith2004} efficiently polarises with a projected polarisation vector being parallel to the symmetry axis. The ring should also contribute in the X-ray range and we model it as an equatorial scattering wedge between 0.03 and 0.05~pc. The symmetry axis of the wedge is aligned with the torus axis and its half-opening angle with respect to the equatorial plane is $20\degr$. The equatorial scattering region is assumed to be at least partly ionised and we therefore assume pure electron scattering with a Thomson optical depth of unity between the inner and outer radius of the wedge.

When showing modelling results we take into account existing (anti-)symmetries between the two hemispheres above and below the equatorial plane. We therefore only discuss the results for viewing directions $0\degr < i < 90\degr$ (or $1 > \cos{i} > 0$) with $i$ being measured with respect to the symmetry axis of the disk and the torus.

\section{X-ray polarisation results}
\label{sec:results}

\subsection{The reprocessing spectrum of the irradiated accretion disc}
\label{subsec:irrad_disc}

\begin{figure}
  \centering
  \includegraphics[width=8.2cm]{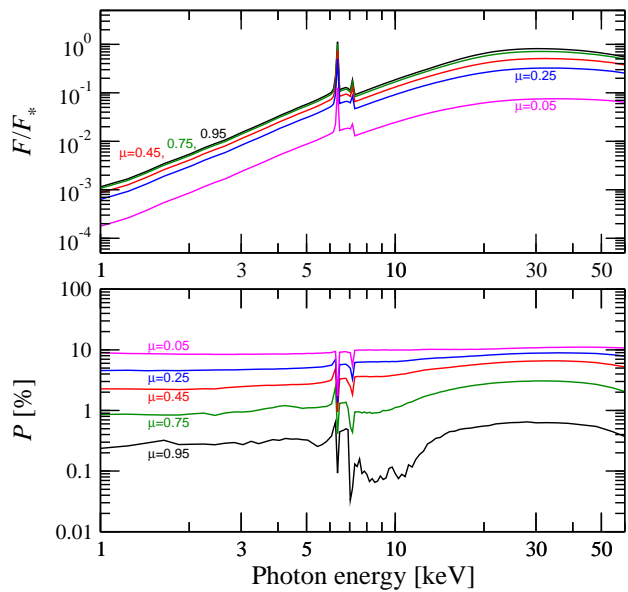}
  \caption{Modelling results for the reprocessed spectrum of an irradiated, neutral accretion disc. Top: spectral flux, $F$, normalised to the centrally emitted flux, $F_{*}$, as a function of the disc inclination $i$ measured by $\mu = \cos{i}$. Bottom: polarisation degree, $P$. The polarisation vector, $\psi$, is aligned with the projected disc axis at all inclinations ($\psi = 90\degr$).}
  \label{fig:disc}
\end{figure}

The spectral modelling of the accretion disc irradiated by an elevated X-ray source confirms previous results, as is shown in Fig.~\ref{fig:disc}. The reprocessed spectra reveal iron K$\alpha$ and K$\beta$ fluorescence lines, the associated iron K absorption edge and the broad Comptonised hump centred around 30~keV. Absorption effects mainly due to hydrogen and helium produce an overall positive slope of the normalised reprocessed spectrum. Note that in the figure we have omitted the contribution of the directly visible, unpolarised primary radiation to have spectral features in the flux and polarisation spectrum come out more clearly.

In this modelling case, the polarisation vector, $\psi$, is always aligned with the projected symmetry axis. At face-on view, when $i$ is low, $P$ does not exceed 1 per cent because the scattering geometry is almost symmetric. The polarisation degree then rises with increasing $i$ and the scattering medium appears less symmetric with respect to the line-of-sight. The polarisation degree drops sharply across the iron K lines, which is due to dilution by the unpolarised fluorescent emission. Across the Compton hump the relation between $P$ and $i$ differs from the soft X-ray band, which is due to the Compton scattering phase function that favours forward over backward scattering at higher photon energies.

\begin{figure}
  \centering
  \includegraphics[width=8.2cm]{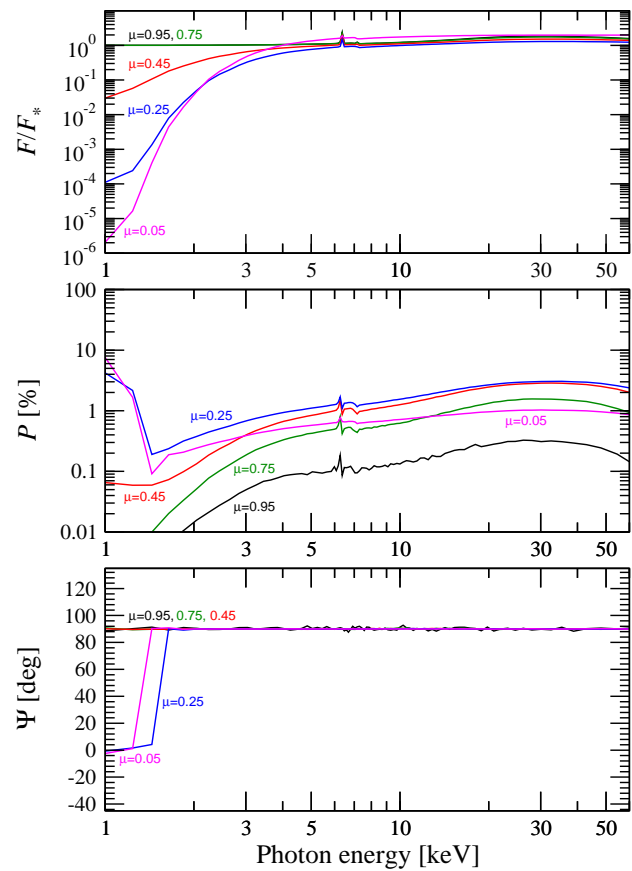}
  \caption{Modelling results for a system of an irradiated accretion disc and an equatorial torus (see Table~\ref{tab:models}) with $N_{\rm H,tor} = 10^{23} {\rm cm}^{-2}$. From top to bottom the figure shows the spectral flux in terms of $F / F_{*}$, the polarisation percentage $P$ and the polarisation position angle $\psi$ as a function of the observer's inclination.}
  \label{fig:torus1e23}
\end{figure}

It is instructive to compare these results to the non-relativistic calculations by \citet{matt1993}. Qualitatively, we find the same behaviour of the polarisation degree and angle but the absolute values of $P$ obtained here are lower than for comparable cases shown in fig.~4 of \cite{matt1993}. This can be explained when considering that the net polarisation degree of the reprocessed radiation results from integrating the reprocessed Stokes flux over the whole disc. After one scattering event, the polarisation of the outgoing radiation is directed perpendicularly to the scattering plane, and thus the central parts of the disc produce polarisation angles around $\psi = 0\degr$ (perpendicular to the symmetry axis) while the outer regions of the disc rather give rise to $\psi = 90\degr$ (parallel to the projected symmetry axis). The net Stokes flux is dominated by the outer parts of the disc but still influenced also by the perpendicular polarisation state coming from the disc centre. Since \cite{matt1993} included a central hole in the disc, the impact of the disc centre on the polarisation was smaller and the net polarisation percentages found were larger.

\begin{figure*}
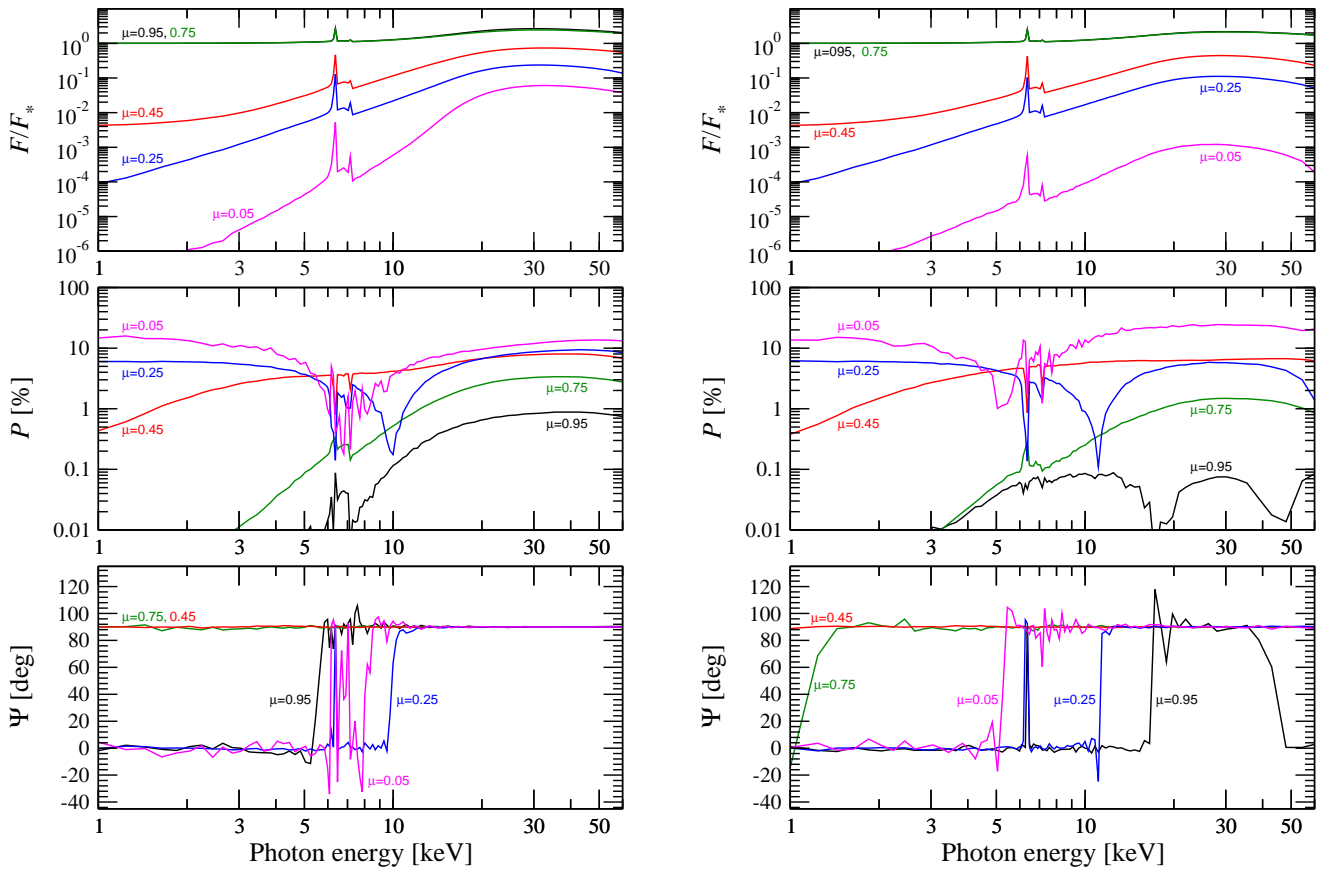

  \centering
  \includegraphics[width=8.2cm]{disk_torus1e25_no_cones.eps}
  \hspace{0.8cm}
  \includegraphics[width=8.2cm]{disk_torus1e27_no_cones.eps}
  \caption{Modelling results for a system of an irradiated accretion disc and an equatorial torus (see Table~\ref{tab:models}) with $N_{\rm H,tor} = 10^{25} {\rm cm}^{-2}$ (left) and $N_{\rm H,tor} = 10^{27} {\rm cm}^{-2}$ (right). The denomination of the panels is as in Fig.~\ref{fig:torus1e23}.}
  \label{fig:torus1e25_1e27}
\end{figure*}

Detailed calculations of the reprocessing by an irradiated accretion disk may also consider the presence of the marginally stable orbit and relativistic ray-tracing effects around the black hole. Such results have recently been presented by \citet{schnittman2010} and \citet{dovciak2011}. The treatment by \citet{schnittman2010} even involves secondary reprocessing, as the reflected component can be directed back to the accretion disc by relativistic light-bending. Note that in the work presented here we assume that the source is located high enough above the disc so that the lack of reprocessing due to the central hole as well as relativistic effects can be neglected.

\subsection{Reprocessing by the torus}

The resulting normalised flux and polarisation spectra for the combined reprocessing by an irradiated accretion disc and an equatorial obscuring torus are shown in Figs.~3--4 for a torus hydrogen column density $N_{\rm H,tor} = 10^{23} {\rm cm}^{-2}$, $N_{\rm H,tor} = 10^{25} {\rm cm}^{-2}$ and $N_{\rm H,tor} = 10^{27} {\rm cm}^{-2}$, respectively. The flux spectra reproduce the features of neutral reprocessing but differ significantly above and below the torus horizon: for $\cos{i} > 0.5$ the irradiating primary source dominates the resulting spectrum. The underlying power-law with $\alpha = 1$ appears as a horizontal curve in the $F/F_*$ representation and the additional reprocessing features are relatively weak. For $\cos{i} < 0.5$, the primary source is no longer directly visible and soft X-ray absorption becomes more pronounced as $N_{\rm H,tor}$ increases. The amount of absorption increases with $i$ as the radiation must undergo more scattering events to escape from the torus funnel toward high inclinations. 

The behaviour of the polarisation degree and angle varies strongly with $N_{\rm H,tor}$. As was pointed out by \citet{ghisellini1994} for a different torus geometry, the net Stokes flux is determined by both reflected and transmitted components that differ in their direction of the polarisation angle. For the case of $N_{\rm H,tor} = 10^{23} {\rm cm}^{-2}$ absorption effects are rather small and mostly apparent in the soft X-ray spectrum at the highest inclinations. Otherwise, the polarisation is determined by transmitted radiation penetrating through the torus walls. The intrinsic polarisation with $\psi = 90\degr$ coming from the accretion disc is thereby reinforced because of scattering in the rather flat geometry of the torus, which has a considerably large half-opening angle of $60\degr$.

When absorption effects become important, then the reprocessed radiation is more dominated by scattering off the torus' inner walls. Since the absorption cross-sections are a strong function of photon energy, the polarisation position angle can switch between the two orthogonal states $\psi = 0\degr$ and $\psi = 90\degr$. Beyond $\sim 12$~keV, where the integrated absorption effects become less important than the efficiency of electron scattering, the hard X-ray polarisation angle tends always to be parallel to the projected torus axis. When going through the switch in polarisation angle, $P$ is near to zero. The polarisation degree also diminishes across the iron K lines due to strong dilution by the intrinsically unpolarised line flux. 

The polarisation degree and position angle depend strongly on the inclination. For a pole-on view, the system of the irradiated disc and torus is close to being axisymmetric and the resulting polarisation of the reprocessed radiation is low. Additionally, the direct unpolarised flux from the primary source further lowers the observed polarisation. When the torus is sufficiently optically thick, as in the case of $N_{\rm H,tor} = 10^{27} {\rm cm}^{-2}$, the net polarisation state is mainly determined by scattering off the inner surfaces of the torus funnel. For a given line-of-sight towards the torus, the relative sizes of its visible inner surfaces oriented perpendicular and parallel to the line-of-sight determine the resulting polarisation angle. Scattering off surfaces which are parallel to the line-of-sight induces aligned polarisation with $\psi = 90\degr$, the perpendicular surfaces, on the other hand, produce perpendicular polarisation with $\psi = 0\degr$. When increasing the viewing angle $i$, the reflected radiation coming from perpendicular surfaces is mainly due to forward and backward scattering, which leads to a weaker Stokes flux for the perpendicular polarisation component. At the same time, the relative exposure of parallel and perpendicular surfaces with respect to the line-of-sight also changes with $i$. The interplay between both effects can be complicated as is witnessed by the partly complex spectral shapes of $\psi$ shown in Figs.~3--4. The explanations given here are similar but not equivalent to the modelling results obtained in the optical/UV band and discussed by \citet{kartje1995} and \citet{goosmann2007}. The differences to the discussion of the optical/UV range are a result of the particular energy dependence of the absorption and scattering cross-sections as well as of the scattering phase functions in the X-ray band.

\subsection{Reprocessing by the disc, a thick equatorial torus and polar cones}
\label{sec:disc_tor_cones}

\begin{figure*}
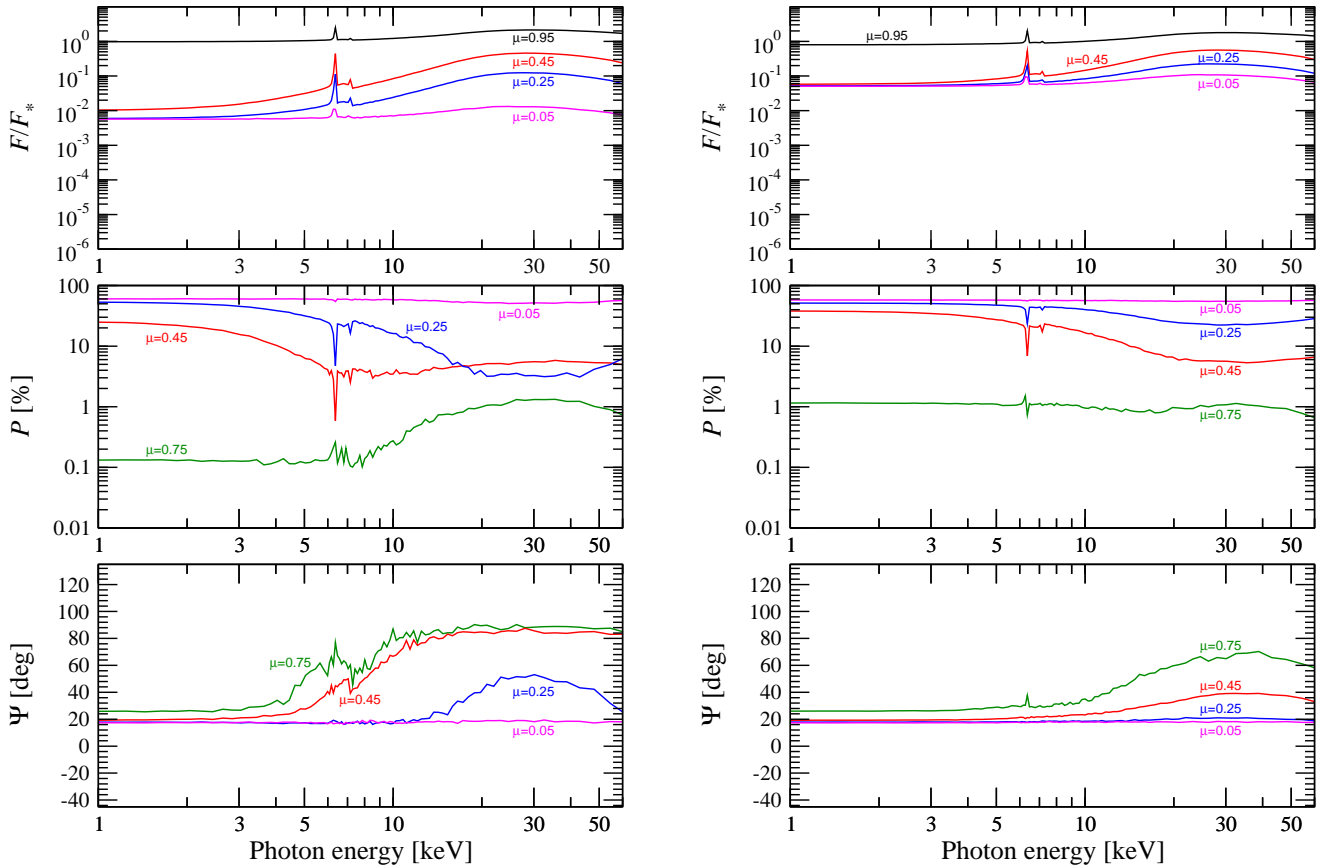

  \centering
  \includegraphics[width=8.2cm]{disk_torus1e27_cones003.eps}
  \hspace{0.8cm}
  \includegraphics[width=8.2cm]{disk_torus1e27_cones03.eps}
  \caption{Modelling results for a system of an irradiated accretion disc, an equatorial torus with $N_{\rm H,tor} = 10^{27} {\rm cm}^{-2}$ and inclined polar electron scattering cones with $\tau_{\rm cone} = 0.03$ (left) and $\tau_{\rm cone} = 0.3$ (right). See Table~\ref{tab:models} for model details. The denomination of the panels is as in Fig.~\ref{fig:torus1e23}.}
  \label{fig:torus_cones003_03}
\end{figure*}

\begin{table*}
 \begin{minipage}{\textwidth}
  \centering
  \caption{Rotation $\Delta \psi = \psi({\rm 20~keV}) - \psi({\rm 2~keV})$ of the polarisation angle and range of the polarisation percentage $P$ between 20~keV and 2~keV for the two values of $\tau_{\rm cone}$ examined in Sect.~\ref{sec:disc_tor_cones} (column 1 and 2) and Sect.~\ref{sec:equatscat} (column 3 and 4).} \label{tab:dpsi}
   \begin{tabular}{@{}lllll@{}}
    \hline
     viewing angle $i$                     & $\tau_{\rm cone} = 0.03$                 & $\tau_{\rm cone} = 0.3$                 & $\tau_{\rm cone} = 0.03$ (equat. wedge) & $\tau_{\rm cone} = 0.3$ (equat. wedge)\\
                     ~                     & $\Delta \psi$, $\Delta P$               & $\Delta \psi$, $\Delta P$              & $\Delta \psi$, $\Delta P$             & $\Delta \psi$, $\Delta P$\\
     \hline                                                                                                                                                             
     $\cos{i} = 0.45$, $i \approx 63\degr$ & $84.9\degr - 20.0\degr = 64.9\degr$     & $32.9\degr - 19.3\degr = 13.6\degr$    & $85.9\degr - 78.4\degr = 7.5\degr$    & $39.9\degr - 27.3\degr = 12.6\degr$\\
                     ~                     & 4.8\% -- 21.3\%                         & 6.8\% -- 36.7\%                        & 7.7\% -- 6.9\%                        & 7.2\% -- 15.9\%\\[3pt]
     $\cos{i} = 0.35$, $i \approx 70\degr$ & $74.6\degr - 18.3\degr = 56.3\degr$     & $23.4\degr - 18.4\degr = 5.0\degr$     & $75.8\degr - 20.1\degr = 55.7\degr$   & $24.7\degr - 20.0\degr = 4.7\degr$\\
                     ~                     & 2.8\% -- 41.3\%                         & 13.6\% -- 45.6\%                       & 4.5\% -- 41.3\%                       & 15.8\% -- 44.6\%\\[3pt]
     $\cos{i} = 0.25$, $i \approx 76\degr$ & $40.6\degr - 17.8\degr = 22.8\degr$     & $20.3\degr - 17.9\degr = 2.4\degr$     & $47.8\degr - 19.3\degr = 28.5\degr$   & $20.9\degr - 19.4\degr = 1.5\degr$ \\
                     ~                     & 4.0\% -- 50.3\%                         & 25.8\% -- 51.0\%                       & 4.4\% -- 48.8\%                       & 26.4\% -- 49.6\%\\[3pt]
     $\cos{i} = 0.15$, $i \approx 81\degr$ & $23.4\degr - 17.5\degr = 5.9\degr$      & $18.6\degr - 17.6\degr = 1.0\degr$     & $24.8\degr - 18.8\degr = 6.0\degr$    & $19.2\degr - 18.9\degr = 0.3\degr$ \\
                     ~                     & 16.6\% -- 56.9\%                        & 43.5\% -- 55.1\%                       & 17.0\% -- 55.2\%                      & 43.8\% -- 53.7\%\\[3pt]
     $\cos{i} = 0.05$, $i \approx 87\degr$ & $18.7\degr - 17.5\degr = 1.2\degr$      & $17.9\degr - 17.3\degr = 0.6\degr$     & $18.8\degr - 18.5\degr = 0.3\degr$    & $18.4\degr - 18.6\degr = -0.2\degr$ \\
                     ~                     & 52.7\% -- 60.2\%                        & 55.3\% -- 57.7\%                       & 53.1\% -- 58.8\%                      & 55.3\% -- 56.4\%\\
    \hline
  \end{tabular}
 \end{minipage}
\end{table*}

To include the effects of polar scattering, we add double-conical electron scattering regions with low optical depth to the previous model of an irradiated accretion disc and equatorial torus. We evaluate the most opaque of our torus models. This choice is motivated by the X-ray spectroscopy data of NGC~1068 confirming that the hydrogen column density deduced from the 0.2--10~keV band must be larger than $10^{25} {\rm cm}^{-2}$ (Matt et al. 1997; Bianchi, Matt \& Iwasawa 2001). Depending on the observer's inclination, the hydrogen column density integrated along the line-of-sight can be by a factor of a few lower than the column density $N_{\rm H,tor}$ that is defined along the equatorial plane. By setting $N_{\rm H,tor} = 10^{27} {\rm cm}^{-2}$ we make sure that the torus remains sufficiently optically thick at all inclinations above $\theta_{\rm tor} = 60\degr$.

We take into account that the double-cone is inclined by $18\degr$ with respect to the torus symmetry axis. We assume a Cartesian coordinate system with the torus symmetry axis being the $z$-axis and the $xy$-plane being the torus equatorial plane. Then, the double-cone symmetry axis is tilted inside the $yz$-plane leaning toward the positive $y$-axis. The angle $\theta_{\rm cone}$ is measured between the double-cone axis and the $z$-axis. Two different Thomson optical depths, $\tau_{\rm cone} = 0.03$ and $\tau_{\rm cone} = 0.3$ are considered.

The results are plotted in Fig.~\ref{fig:torus_cones003_03}. For each inclination, we consider the azimuthal viewing angle, $\phi$, along the positive $x$-axis integrating the flux over $20\degr$ in azimuth. Such a view corresponds to the illustration given in Fig.~\ref{fig:sketch}. Compared to the models without polar scattering, the soft X-ray spectrum seen at $i > \theta_{\rm tor}$ is much less absorbed. Even at low optical depth, the double-cone scatters an important fraction of the primary emission around the torus towards high inclinations. This is the `periscope effect' of polar scattering that has been evaluated and observationally exploited in the optical/UV range \citep{antonucci1985,smith2002}. Since the periscope effect is due to polar electron scattering, it is not surprising that it also exists in the X-ray range. Even at low $\tau_{\rm cone}$ the continuum polarisation vector in the soft X-ray band is dominated by polar scattering inducing a polarisation angle of $\psi = 18\degr$ that corresponds to the inclination of the outflows. The soft X-ray polarisation rises with $i$ reaching a maximum of about 60 per cent at edge-on view. 

At higher photon energies, the polarisation induced by the torus becomes more important because its albedo rises. The polarisation spectrum is again highly influenced by dilution across the iron K lines, but past the iron K absorption edge the Compton scattering off the torus becomes dominant for intermediate inclinations. The exact outcome of the polarisation depends on the relative Stokes fluxes coming from the polar outflows, on the one hand side, and from the accretion disc and torus, on the other hand side. When viewed strictly edge-on the absorption effects in the torus are too strong and the polarisation angle is always determined by the polar outflows. But a rotation of the polarisation angle between the soft and the hard X-ray band is observed and could be used to measure the misalignment between the symmetry axes of the torus and the outflows. In the first and second column of Table~\ref{tab:dpsi}, we quantify the difference $\Delta \psi$ in the polarisation angles between 20~keV and 2~keV for different obscured viewing angles. The values of $\Delta \psi$ vary strongly with the inclination but even at edge-on view and for a high optical depth of the ionisation cone a small rotation of the polarisation angle is noted.

\begin{figure}
  \centering
  \includegraphics[width=8.2cm]{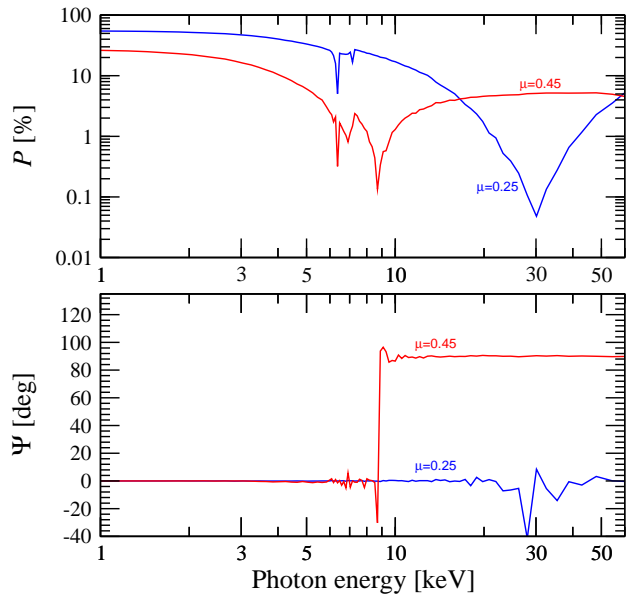}
  \caption{Polarisation percentage and position angle as a function of photon energy for a model of an irradiated accretion disc, an equatorial torus with $N_{\rm H,tor} = 10^{27} {\rm cm}^{-2}$ and polar scattering cones with $\tau_{\rm cone} = 0.03$ seen at the obscured viewing angles $\mu = 0.45$ and $\mu = 0.25$. In this case, the polar outflows are aligned to the $z$-axis and the results can be contrasted to the case of Fig.~\ref{fig:torus_cones003_03}~(left).}
  \label{fig:torus_cones003_noinc}
\end{figure}

For contrast, we show in Fig.~\ref{fig:torus_cones003_noinc} the results for the X-ray polarisation of a model including polar outflows with $\tau_{\rm cone} = 0.03$ but being aligned to the $z$-axis. Without inclining the double-cone, $\psi$ can only adopt the two possible values that correspond to the perpendicular polarisation caused by the polar scattering in the soft X-ray band and the aligned polarisation vector due to the disc/torus at higher photon energies. This bimodal distribution of the polarisation vector is related to the symmetry of the model with respect to the $z$-axis. The comparison shows that the gradual change of $\psi$ with photon energy, as it is apparent in the modelling case of Fig.~\ref{fig:torus_cones003_03}~(left), is characteristic for a misalignment of the outflows with respect to the torus. 

\subsection{The effect of equatorial scattering}
\label{sec:equatscat}

The third and fourth column of Table~\ref{tab:dpsi} show some exemplary results for a scenario that includes equatorial scattering. The corresponding models are based on the cases of Sect.~\ref{sec:disc_tor_cones} but additionally include an equatorial wedge as described in Table~\ref{tab:models}. The additional scattering region has an impact at all viewing directions. For polar outflows with low optical depth $\tau_{\rm cone}$, the rotation $\Delta \psi$ of the polarisation angle can either rise or fall with respect to the models of Sect.~\ref{sec:disc_tor_cones}; for a high $\tau_{\rm cone}$, the value of $\Delta \psi$ always diminishes. At edge-on inclination, the polarised flux escaping from the torus funnel is too small to efficiently counterbalance the effect of polar scattering. Therefore, the resulting $\Delta \psi$ is very low.

The results illustrate that the wavelength-dependence of the scattering efficiency and the radiative coupling between the different reprocessing regions are important to understand the net polarisation. The equatorial electron scattering competes with the polar electron scattering in the outflows, independently of spectral energy. Therefore, the polarisation position angles are systematically higher when equatorial scattering is included. The rise in $\psi$ is more apparent at intermediate inclinations but still exists towards an edge-on view.

The additional scattering region also has an impact on the rotation $\Delta \psi$ between the soft and hard X-ray polarisation angle. The equatorial wedge covers a fraction of the primary radiation going towards the inner surfaces of the torus and it changes the irradiation geometry for both the torus and the polar cones. These effects diminish the resulting $\Delta \psi$, at least when the optical depth in the polar outflows is high. We have investigated different half-opening angles of the pure electron scattering wedge as well as a reduction of its optical depth. The obtained values of $\Delta \psi$ change for such cases, but the qualitative behaviour stays the same.

\section{Discussion}
\label{sec:discuss}

\subsection{Directly measuring the orientation of the inner outflows}
\label{sec:discuss_orient}

We have conducted accurate modelling of the expected X-ray polarisation induced by complex reprocessing in the active nucleus of NGC~1068. This work is motivated by the apparent misalignment of the ionisation cones with respect to the torus axis as discussed by \citet{raban2009}. The orientation of the ionisation cones was inferred from slit spectroscopy of the narrow line region (NLR) using the {\it HST STIS} in combination with kinetic modelling \citep{das2006}. But \citet{raban2009} also point out that this orientation is still a matter of debate. From the observation of infrared emission lines, Poncelet, Sol \& Perrin (2008) deduce a different position angle of the NLR that is more in agreement with the {\it HST} imaging results of [OIII] emission reported by \citet{evans1991}. But then the UV-imaging probably suffers from significant foreground absorption. In addition to that, we want to point out that all these measurements are taken on large spatial scales reaching out to 100~parsec from the central engine. It is not straightforward to derive from these observations what the actual geometry of the ionisation cones across a few parsec is. One could imagine that the outflows become deflected or twisted on intermediate scales and thus their orientation would vary with distance. As \citet{raban2009} conclude, further investigation is needed to really constrain the geometry of the innermost outflows.

In this context, X-ray polarimetry is going to give unambiguous constraints. The scattered X-ray emission must come from very close to the central engine, as is indicated from X-ray spectroscopy results revealing the presence of high-ionisation emission lines of iron \citep{marshall1993,matt2004}. Furthermore, any foreground absorption effects seen in the UV should much less interfere in the X-ray range so that X-ray polarimetry enables a direct view on the innermost parts of the outflow. Our modelling results have shown that the polarisation position angle in the soft X-ray range is clearly dominated by the polar scattering, which enables a precise measurement of the orientation of the scattering material.

\subsection{Determining the misalignment with the torus}

Another aim of this investigation is to verify if the rotation $\Delta \psi$ of the X-ray polarisation angle between soft and hard X-ray energies can be used to measure the misalignment between the polar scattering regions and the obscuring torus. We have found that coherent modelling of the different scattering components leads to values of $\Delta \psi$ that at heavily obscured viewing directions allows one to distinguish the torus from the ionisation cones. The relation between $\Delta \psi$ and the actual misalignment also depends on $i$. The highest values of $\Delta \psi$, which are most favourable for future X-ray polarimetry observations, are found at moderate viewing angles. At strictly edge-on viewing directions, $\Delta \psi$ becomes very small, especially for higher optical depths of the ionisation cones. But, as we discuss in the following, our modelling approach is rather conservative, and in reality we should expect more favourable conditions for the measurement of $\Delta \psi$. The results we present in this work give lower limits on $\Delta \psi$ for NGC~1068 assuming the least favourable conditions for a rotation of the polarisation angle between low and high photon energies.

\subsubsection{Optical depth and geometry of the outflow}
\label{sec:discuss_cones}

The net Stokes flux and therefore the resulting polarisation angle is determined by the competition between the disc/torus system and the polar scattering cones. Especially at higher inclination the polar scattering has a strong impact as the scattering-induced polarisation rises towards orthogonal scattering angles. The efficiency of polar scattering depends on $\tau_{\rm cone}$ and is low for optically thin scattering cones. It rises with increasing $\tau_{\rm cone}$ and then goes through a maximum until multiple scattering effects become important and diminish the resulting polarisation. Note also, that the Stokes flux from the double-cone results from integrating photons over its entire opening angle. This leads to combining scattered photons having a large range of polarisation orientations and also diminishes the net polarised flux. Overall, the case of $\tau_{\rm cone} = 0.3$ that we investigate produces a strong Stokes flux but at most inclinations it remains comparable to the polarised flux coming from the disc/torus system and therefore the rotation $\Delta \psi$ is significant. Having in mind the unified scheme of AGN, it is important that $\tau_{\rm cone}$ remains low enough as the ionised winds are seen in transmission in Seyfert-1 galaxies, where they produce warm absorption in the soft X-ray band. Observations of the warm absorber rule out that the medium has a higher column density than $10^{23} {\rm cm}^{-2}$ for most objects \citep[see][and references therein]{turner2009}. Only a few exceptions exceed this threshold so that our choice of $\tau_{\rm cone} = 0.3$, which corresponds to a column density around $4.5 \times 10^{23} {\rm cm}^{-2}$, is a good estimate for the upper limit. Note also that for NGC~1068 the column density of the polar scattering medium has been constrained from the observed emission measure distribution to $4 \times 10^{21} {\rm cm}^{-2} < \log{N_{\rm H,Cone}} < 7 \times 10^{22} {\rm cm}^{-2}$ \citep{kinkhabwala2002}. Furthermore, we assume in our modelling that the matter is uniformly distributed inside the double-cone, whereas the approach by \citet{das2006} is based on outflows with a shell-structure, such as predicted for magnetically driven outflows \citep{konigl1994}. For all these reasons, the Stokes flux coming from the ionisation cones of NGC~1068 should be lower than we assume in our modelling, which enables a larger $\Delta \psi$. 

The misalignment of the ionisation cones with respect to the torus axis as derived by \citet{raban2009} refers to the projection of the double-cone onto the plane of the sky. But if the outflows are rotated in azimuth with respect to the $yz$-plane, it is possible that the real misalignment with respect to the torus axis is larger than $18\degr$. Note that a slight inclination of the double-cone towards the observer is suggested by the kinematic modelling of \citet{das2006}. Then, the average scattering angle for the polar electron scattering is no longer close to $90\degr$ but will shift towards forward and backward scattering. This reduces the polarised flux coming from the outflows. At the same time, a stronger inclination of the cones also requires a larger half-opening of the torus, otherwise the outflows would be blocked. The scattering distribution of the torus is thus geometrically flatter in this case, which should increase its Stokes flux. Both effects work in favour of higher values for $\Delta \psi$ than found from the models presented here.

\subsubsection{Clumpiness of the torus material}
\label{sec:discuss_torus}

\begin{figure}
  \centering
  \includegraphics[width=8.2cm]{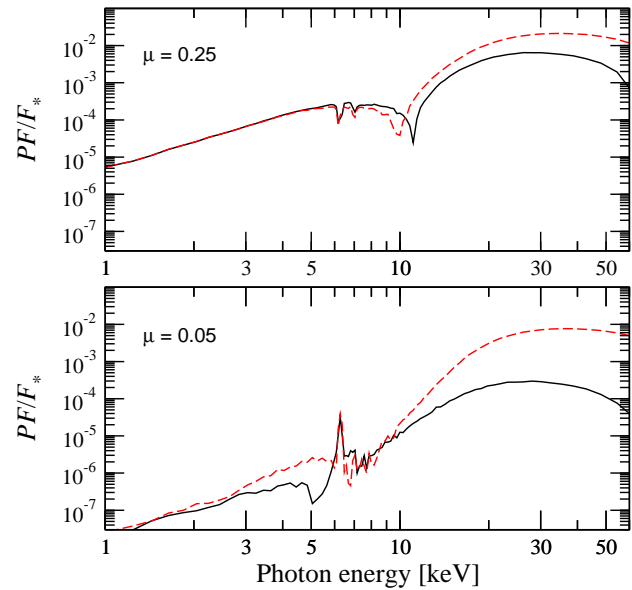}
  \caption{Modelled polarised flux spectra for a system of an irradiated accretion disc and an equatorial torus with $N_{\rm H,tor} = 10^{27} {\rm cm}^{-2}$ (solid line) and $N_{\rm H,tor} = 10^{25} {\rm cm}^{-2}$ (dashed line) for two different obscured viewing angles.}
  \label{fig:torPF}
\end{figure}

We are also pessimistic in our assumptions about the torus scattering properties. In our modelling we assume a maximum optical depth by setting $N_{\rm H,tor} = 10^{27} {\rm cm}^{-2}$. This agrees with previous spectroscopic studies showing that the obscuring material along the line-of-sight toward NGC~1068 is entirely opaque. But for most viewing angles we are still more than one order above the lower threshold of the observed hydrogen column density, which was given by \citet{matt1997} and \citet{bianchi2001} based on {\it ASCA} and {\it BeppoSAX} data. It is thus likely that the torus is less optically thick than we assumed in our modelling. A lower $N_{\rm H,tor}$ increases the polarised flux of the torus as we show in Fig.~\ref{fig:torPF}. Between the cases of lower and higher $N_{\rm H,tor}$ there is an increase in Stokes flux at 20~keV by a factor of 2.5 at $i \approx 76\degr$ and of more than a factor of 10 at $i \approx 87\degr$. 

The exact geometry of the obscuring dust region remains a matter of debate, although infrared interferometry observations start putting robust constraints on the geometrical size of the dusty region \citep{jaffe2004,ramos2009,hoenig2010a}. The latest generation of infrared radiative transfer models allows for a satisfying description of the data by assuming that the dust region is clumpy \citep{nenkova2008a,nenkova2008b,schartmann2008,hoenig2010b}. Such a clumpy composition of the torus could also address criticism that the unified model could not properly explain how the observed spatial extension of the dust region can be maintained against self-gravity. The clumpy structure would argue for a dynamical model in which the torus is not in hydrostatic equilibrium \citep{krolik1988}. In the results presented here we do not include the effect of clumpiness on the resulting polarisation. But if the torus medium is clumpy and at the same time dynamical, there is a better chance for fluctuations in the column density along a given line-of-sight. Observing NGC~1068 with a lower opacity $N_{\rm H,tor}$ is thus more likely if the medium is clumpy instead of uniform. As shown above, a lower opacity enhances the Stokes flux from the torus and thus the rotation $\Delta \psi$.

\subsubsection{The primary source and equatorial scattering}
\label{sec:discuss_primary}

We assume that the polarisation coming from the accretion disc is based on reprocessing across the whole disk surface. The extended disk has a large radius compared to the height of the X-ray source above. Then, the outer disc regions produce polarisation with $\psi = 90\degr$. This aligned polarisation is being diminished by the Stokes flux coming from the disk centre and having $\psi = 0\degr$. As discussed in Sect.~\ref{subsec:irrad_disc}, such a setup assumes that the primary source is located far above the black hole so that the central hole of the disk does not have any significant impact. When relaxing this assumption and considering sources that are much closer to the black hole, the reprocessing geometry changes \citep[see e.g.][]{martocchia1996,miniutti2004,schnittman2010,dovciak2011} and the missing part of the disc inside the marginally stable orbit becomes significant for the Stokes flux. The net polarisation with $\psi = 90\degr$ of the disc is thus stronger for such a case, which works in favour of a larger value for $\Delta \psi$. Assuming a different irradiation geometry, for instance a central but more extended hot corona, would not change this result much as is implied by the results of \citet{matt1993}. The main reason for this is that the disc remains very large with respect to the size of the corona.

We have shown that generally equatorial scattering leads to a lower $\Delta \psi$, in particular at high inclinations of the observer. But note that we assume the equatorial scattering being entirely due to electrons. This is an extreme example as beyond the BLR also moderately ionised and neutral medium should exist. Less ionised matter has a different scattering efficiency and should absorb more strongly in the soft X-ray band and thereby maintain higher values of $\Delta \psi$. The two exemplary cases shown in Sect.~\ref{sec:equatscat} rather define lower limits of the expected $\Delta \psi$ when equatorial scattering is important.

\subsection{Encouraging prospects for future X-ray polarimetry of NGC~1068}

The observational window of X-ray polarimetry will soon be (re-)opened by the NASA {\it Gravity and Extreme Magnetism Small Explorer (GEMS)} that is currently prepared for launch in 2014 \citep{swank2009,jahoda2010}. The satellite will be equipped with a soft X-ray polarimeter reaching up to 10~keV. Confirmation of the polarisation position angle induced by scattering in the misaligned outflows of NGC~1068 will thus principally be possible with {\it GEMS}. For the practical measurement, the collecting area of the mirrors might still be a limiting factor. On the other hand side, {\it GEMS} is entirely devoted to X-ray polarimetry and allows for long exposure times.

The technology is also ready for mid-size observatories that include imaging polarimetry at high angular resolution and over the whole range of 2~keV to 35~keV. Such X-ray polarimeters could efficiently observe near-by AGN. We use the response matrices of the polarimeters designed for the formerly proposed {\it New Hard X-ray Mission} \citep{tagliaferri2010,soffitta2010}. It turns out that a 300~ks observation of NGC~1068 would be sufficient to constrain the polarisation angle with a statistical uncertainty of $5\degr$ if the polarisation degree is 10 per cent across 6~keV to 35~keV. For higher values of $P$, the precision on the measurement rises; a polarisation of 20 per cent constrains the angle with an uncertainty of $2.5\degr$. At lower photon energies, the performance of the X-ray polarimeter is expected to be even better. The soft X-ray polarisation obtained by our modelling is at least comparable to and often significantly higher than 20 per cent as shown in Table~\ref{tab:dpsi}. Therefore, a future mid-size X-ray observatory with broad band polarimetry capabilities could directly measure the orientation of the innermost ionisation cones as described in Sect.~\ref{sec:discuss_orient}.
 
If we assume a viewing angle of $i = 70\degr$ towards NGC~1068, as found by \citet{hoenig2007} from applying a clumpy torus radiative transfer model to infrared data, then the expected values for $\Delta \psi$ given in Table~\ref{tab:dpsi} are also within the measurable limits. Should the inclination angle be higher as might be suggested by water maser observations \citep{gallimore1996,greenhill1996} then the possibility to measure $\Delta \psi$ depends on the model details. We consider the values given in Table~\ref{tab:dpsi} as lower limits that are based on a conservative modelling approach.

Note that if one decides to include both, a soft and a hard X-ray polarimeter (up to about 25~keV) in the large mission design of the {\it International X-ray Observatory} \citep{barcons2011}, then measurements of the broad band polarisation still further improve. The method described here might then be applied systematically to a larger sample of objects. In summary, we therefore think that future X-ray polarimetry over a broad energy range is a promising tool to disentangle the details of the inner AGN geometry.

\section*{Acknowledgements}

The authors want to thank Stefano Bianchi for helpful discussion on X-ray spectroscopy results for NGC~1068 and Fabio Muleri for providing estimates of the expected polarimetry results from a mid-size observatory. We are grateful to the anonymous referee for helping us clarifying the paper. Ren{\'e} Goosmann acknowledges support of this work by the French GdR PCHE.

\bsp

\label{lastpage}

\end{document}